\documentclass[twocolumn,showpacs,showkeys,preprintnumbers,amsmath,amssymb,prl]{revtex4}

\usepackage{graphics,color}
\usepackage{epsfig}
\usepackage{pifont}

\begin{document}

\title{Non-vanishing Berry Phase in Chiral Insulators}
             
\author { J{\"u}rgen~K{\"u}bler$^*$$^1$, Binghai Yan$^2$,
 and Claudia Felser$^2$ }
\affiliation{$^1$Institut f{\"u}r Festk{\"o}rperphysik, Technische Universit{\"a}t Darmstadt, D-64289 Darmstadt, Germany}
\affiliation{$^2$Max Planck Institute for Chemical Physics of Solids,
             D-01187 Dresden, Germany}      
             
\email{juergen.kuebler@gmail.com}             

\date{\today}


\begin{abstract}
The binary compounds FeSi, RuSi, and OsSi are chiral insulators crystallizing in the space group $P2_13$ which is cubic. By means of \textit{ab initio} calculations we find for these compounds a non-vanishing electronic Berry phase, the sign of which depends on the handedness of the crystal. There is thus the possibility that the Berry phase signals the existence of a macroscopic electric polarization due to the electrons. We show that this is indeed so if a small external magnetic field is applied in the [111]-direction. The electric polarization is oscillatory in the magnetic field and possesses a signature that distinguishes the handedness of the crystal. Our findings add to the discussion of topological classifications of insulators and are significant for spintronics applications, and  in particular, for a deeper understanding of skyrmions in insulators.

\end{abstract}
\maketitle

A chiral structure is a distribution of atoms that cannot be brought into coincidence with its mirror image by applying any choice of rotations and translations \cite{flack}. To be chiral their space group must not contain inversion, rotation-inversion and mirror planes. Recently the phenomenon of chirality has attracted the attention of condensed matter scientists in the context of magnetic materials with non-centro symmetric structures (skyrmions) and semiconductors with metallic surface states (topological insulators, TI). The metallic surfaces in TI are protected chiral spin currents \cite{zhang}, a result of a special electronics structure in reciprocal space. Skyrmions are topological particles in real space, magnetic screw like nanostructures, which can be directly observed by neutron scattering \cite{muhl} or Lorenz microscopy \cite{yu}. 

 Well-known examples of chiral structures are MnSi and MnGe where the magnetic structure is linked with the chiral crystal structure \cite{grigoriev}.
The latter are cubic, of the type B20, which is described by the space group $P2_{1}3$, international tables for crystallography no. 198. Even more spectacular are magnetic skyrmions observed in the chiral magnets MnSi \cite{ross,muhl}, Fe$_{1-x}$Co$_x$Si \cite{munzer,yu} and FeGe \cite{yu1}. These vortex-like spin structures are of great fundamental interest as well as promising candidates for spintronics applications. Most recently, skyrmions were found in a multiferroic chiral insulator \cite{seki}. The materials discussed here, FeSi, RuSi, and OsSi \cite{wata,bucher,pear}, are non-magnetic insulators crystallizing in the same crystal structure.

 FeSi has attracted a great deal of attention, both experimentally and theoretically. It is an insulator at low temperatures but looses the gap at about 200 K. In the past the central question addressed the origin of the gap and the reason for Fe being nonmagnetic, spawning a number of interesting concepts and theoretical models \cite{mason,fu}. Here we take it as a prototypical example for a chiral insulator where, among other things, the role of symmetry can be studied without any complicating magnetic effects.

\begin{figure}[htb]
   \begin{center}
   \epsfig{file=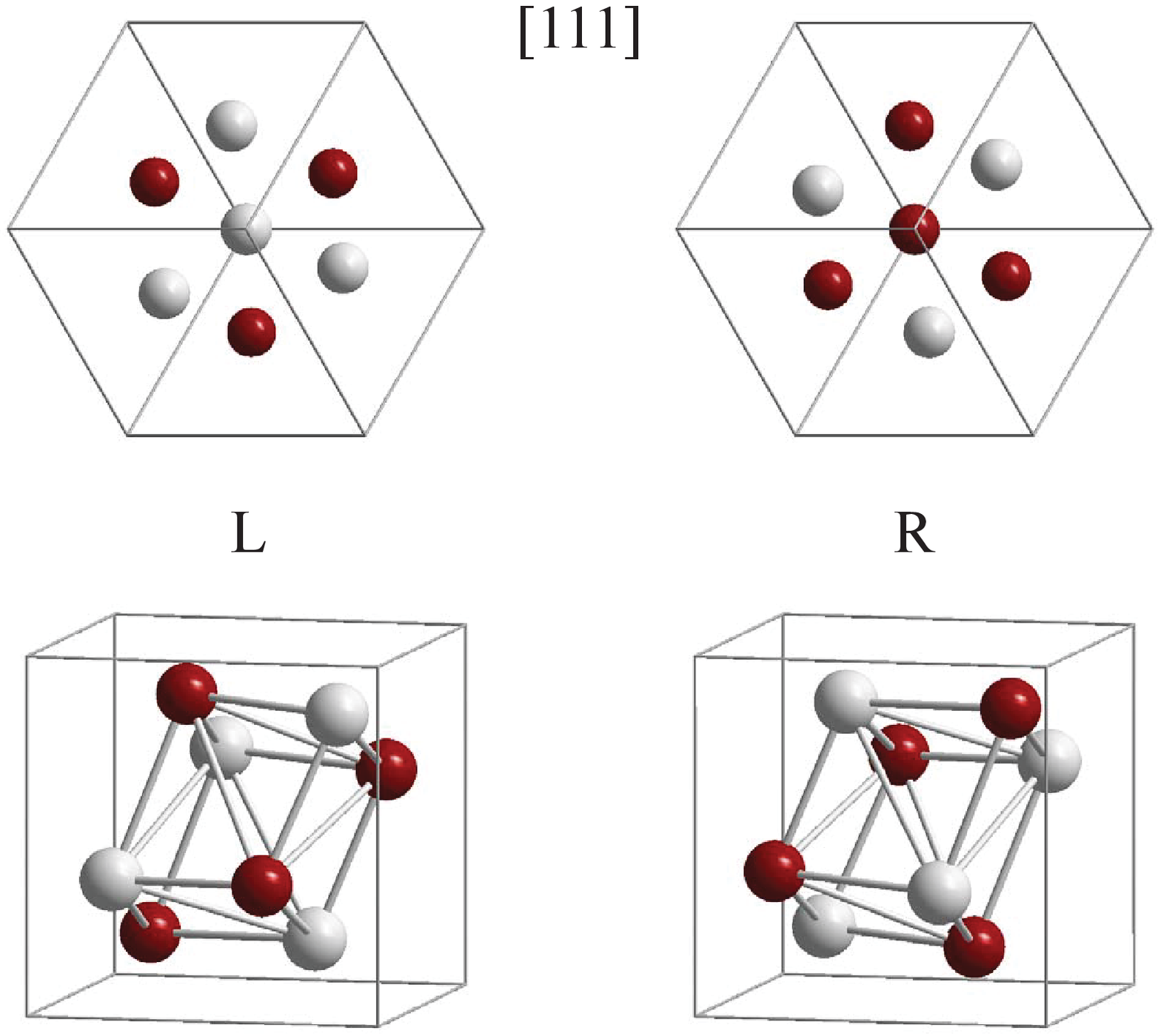,width=7cm} 
   \caption{(Color online) The two chiral crystal structures of FeSi (RuSi, OsSi). 
            Red: Fe (Ru, Os)atoms, grey: Si atoms. 
            The structure on the left is the reported one. The top views are along the [111] axis, the body diagonal of the cube.} 
   \label{fig1}
   \end{center}
\end{figure}

The crystal structure of FeSi (also of RuSi and OsSi) is graphed in Fig.\ref{fig1}. The space group  $P2_13$ is  characterized by two experimental parameters $u(\rm {Fe})= 0.137$ and $u(\rm {Si})=0.842$, Fe and Si both occupying the Wyckoff position 4$a$, FeSi thus having eight atoms per simple-cubic unit cell. This structure has chiral symmetry and occurs in two forms both of which are shown in Fig~\ref{fig1}. The values of $u$ above describe the left hand side of the figure. 
\begin{figure}[htb]
  \begin{center}
  \epsfig{file=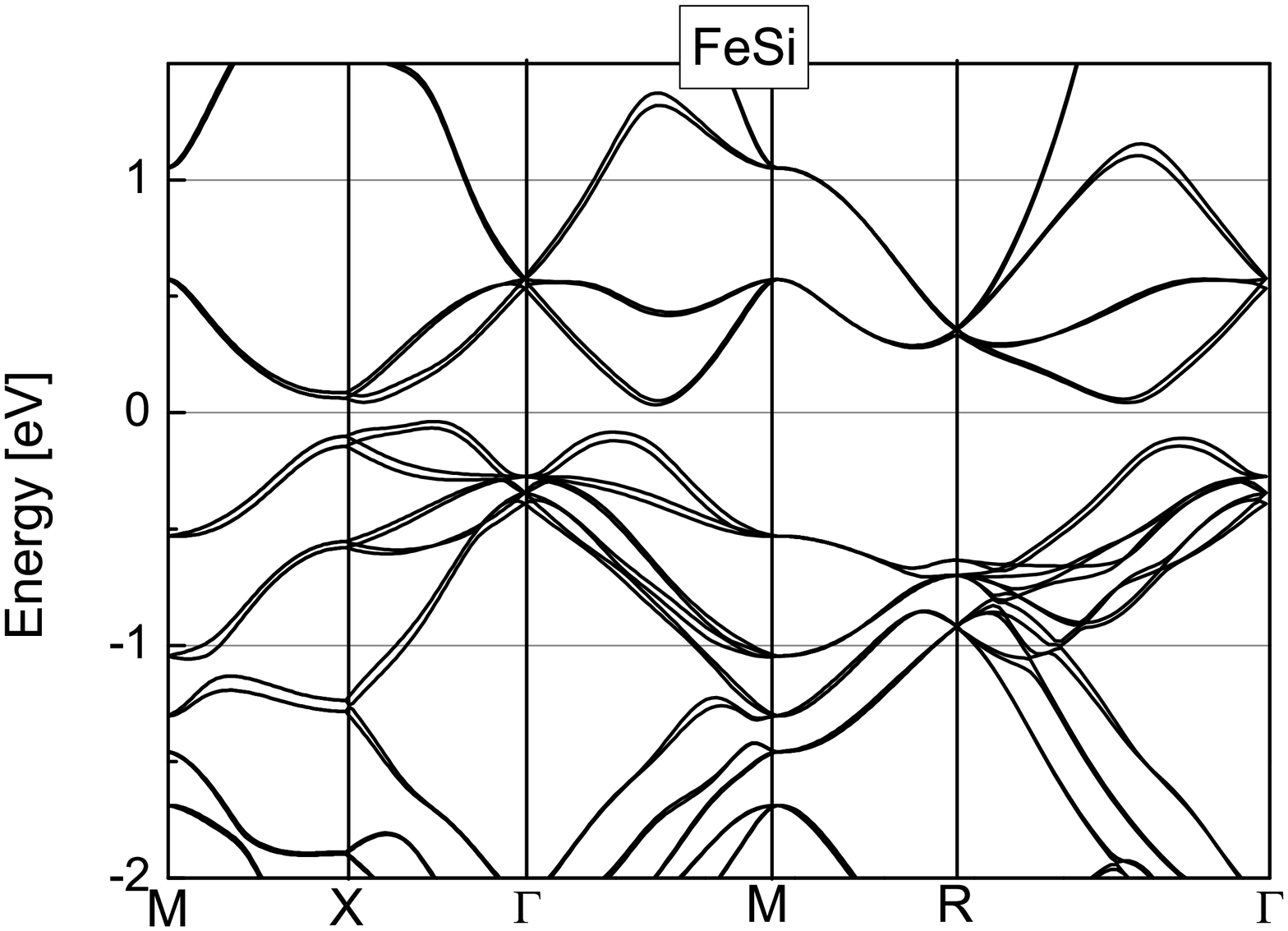,width=6cm} \epsfig{file=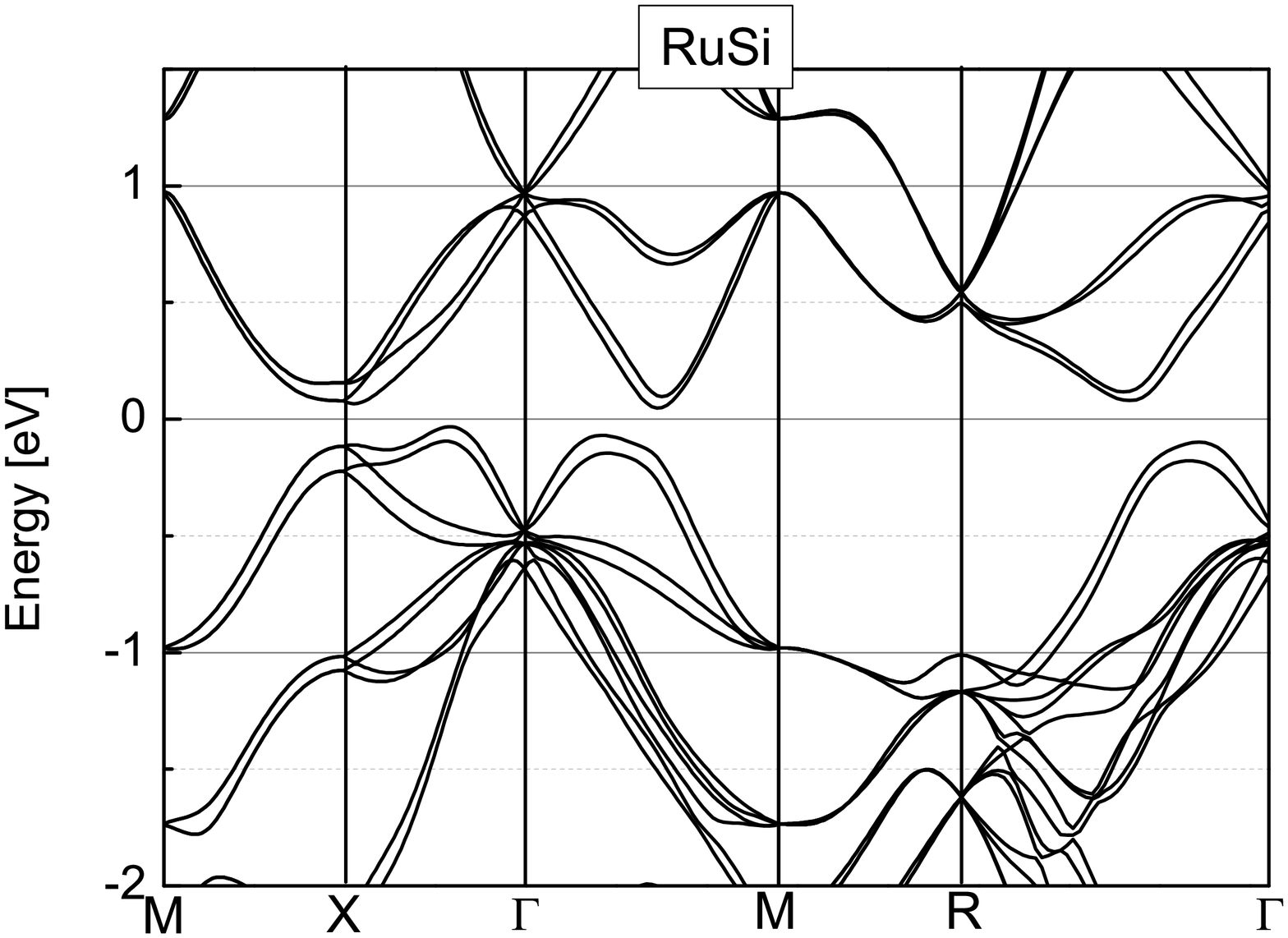,width=6cm}
  \caption{ The band structure of FeSi and RuSi for energies near the band gap which is chosen 
            to be centered at the origin of energy. Spin-orbit coupling is included.  }
  \label{fig2}
  \end{center}
\end{figure}

The crystal structure displayed in the left part of Fig.~\ref{fig1} is described as follows. One Fe atom (red) and one Si atom (grey) on the lower left and upper right, respectively, reside on the body diagonal [111] of the cubic cell. The other atoms, three Fe and three Si atoms form equilateral triangles whose center is cut by the body diagonal which is normal to the triangles, as is seen from the view along the [111] axes (upper part of Fig~\ref{fig1}). Usually the lattice parameters are crystallographically standardized in a way that the transition metal sublattice, here Fe, is lefthanded whereas the main group element sublattice, here Si, is righthanded. This convention is used independent of the "real" chirality of the material. In what follows Fe may be imagined to be replaced by Ru or Os.

The band structure of FeSi has been calculated before, for instance by ref.~\cite{mattheiss,fu}.
The band structure is non-symmorphic, therefore the bands without spin-orbit coupling are 4-fold degenerate at the symmetry lines $M-X$ and $M-R$. When spin-orbit coupling (SOC) is included, much of the 4-fold symmetry is lifted, but Kramers degeneracies remain. The band-structure for FeSi is shown in  Fig.~\ref{fig2} for energies near the band gap. It is to be compared with that of RuSi also shown in  Fig.~\ref{fig2}. The effects of SOC are small in FeSi but are more distinct in RuSi. These results were obtained with the ASW method \cite{kubler} in the local density functional approximation (LDA). As is usually the case in the LDA, the energy gap is underestimated. For the lattice constants and $u$  the experimental values were used \cite{wata,bucher,pear}.

\begin{figure}[htb]
  \begin{center}
  \epsfig{file=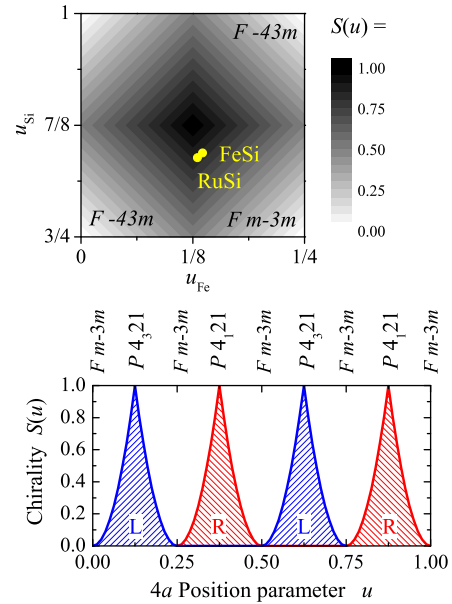,width=6cm} 
  \caption{(Color online) Chirality in space group 198.
           The lower part shows the changes of the symmetry of a single sublattice of the $P2_13$ group as a function of $u$.
           L and R designate left and right handed chirality. Plotted is the continuous chirality measure. 
           The upper part of the figure shows the 2D distribution of the chirality accounting for both, the Fe and the Si sublattices.
           The dots mark the experimental $u$-values of FeSi and RuSi.} 
  \label{fig11}
  \end{center}
\end{figure}

Mattheiss and Hamann \cite{mattheiss} in their discussion of the band structure of FeSi pointed out the connection of FeSi in the $P2_13$ structure with FeSi in the NaCl ($Fm\bar{3}m$) structure, in which FeSi is metallic. We put this in a more general context and show in Fig.~\ref{fig11} the symmetry changes of one sublattice of FeSi as a function of the parameter $u$.  
  
We see, when $u$ increases from 0 to 1, that the $P2_13$ symmetry is interrupted by the chiral symmetry $P4_332$ for $u=0.125$, the achiral symmetry $Fm\bar{3}m$ for $u=0.25$, the chiral symmetry $P4_132$ for $u=0.375$, etc while the handedness of the $P2_13$ structure changes from left (L) to right (R) and back again. The continuous chirality measure~\cite{zabrodsky} is used in Fig.~\ref{fig11} to quantify the chirality of the lattice. It is defined by $S(u) = \sum_{i=1}^n \left\| p_i - p_i^{\rm sym} \right\| ^2$, where $p_i$ are the positions of the Fe and Si atoms in $P2_13$ symmetry and $p_i^{\rm sym}$ are the corresponding positions in achiral symmetry, when the FeSi lattice becomes
NaCl ($Fm\overline{3}m$) or ZnS ($F\overline{4}3m$) type. In Fig.~\ref{fig11}, $S(u)$ is normalized to the maximum that appears when the structure adopts $P4_132$ or $P4_332$ symmetry.   

We now show that the chiral crystal structure $P2_13$ is characterized by a non-vanishing Berry phase \cite{berry,zak} in the electronic system. To do this we must make a decision about the choice of the origin, \textit{i.e.} the location $\textbf{r}_{\tau}$ of the atom representing the sublattice $\tau$. We shift the atoms such that the metal atom on the diagonal is found at the origin of the simple cubic lattice. The occasional problem of differing choices of the origin in the experimental data \cite{pear} is thus removed and the Bravais lattice is defined by the charge of one of the metal atoms \cite{vander}.
The Berry phase (BP) is calculated by evaluating numerically a standard expression given by King-Smith and Vanderbilt \cite{king}. One obtains
 $\phi^{(\lambda)}(\mathbf{k}_{\bot})$, which is the BP for a path in reciprocal space beginning at $\mathbf{k}_{\bot}$ on one side of the Brillouin zone and ending on the opposite side.
The BP is summed up $\Phi^{(\lambda)}=\int_{A}d\mathbf{k}_{\bot}\phi^{(\lambda)}(\mathbf{k}_{\bot})$
where the surface integral is carried out over a discrete mesh. $\Phi^{(\lambda)}$ is a vector component normal to the area $A$. 
 For the simple cubic unit cell $A$ is  taken to be one of the 3 faces of the Brillouin zone cube thus giving one of the three components of the BP vector, which all turn out to be equal (as required by symmetry). The computations are converged with a reasonable number of $k$-points, 3840 for a given area $A$.

\begin{figure}[htb]
   \begin{center}
   \epsfig{file=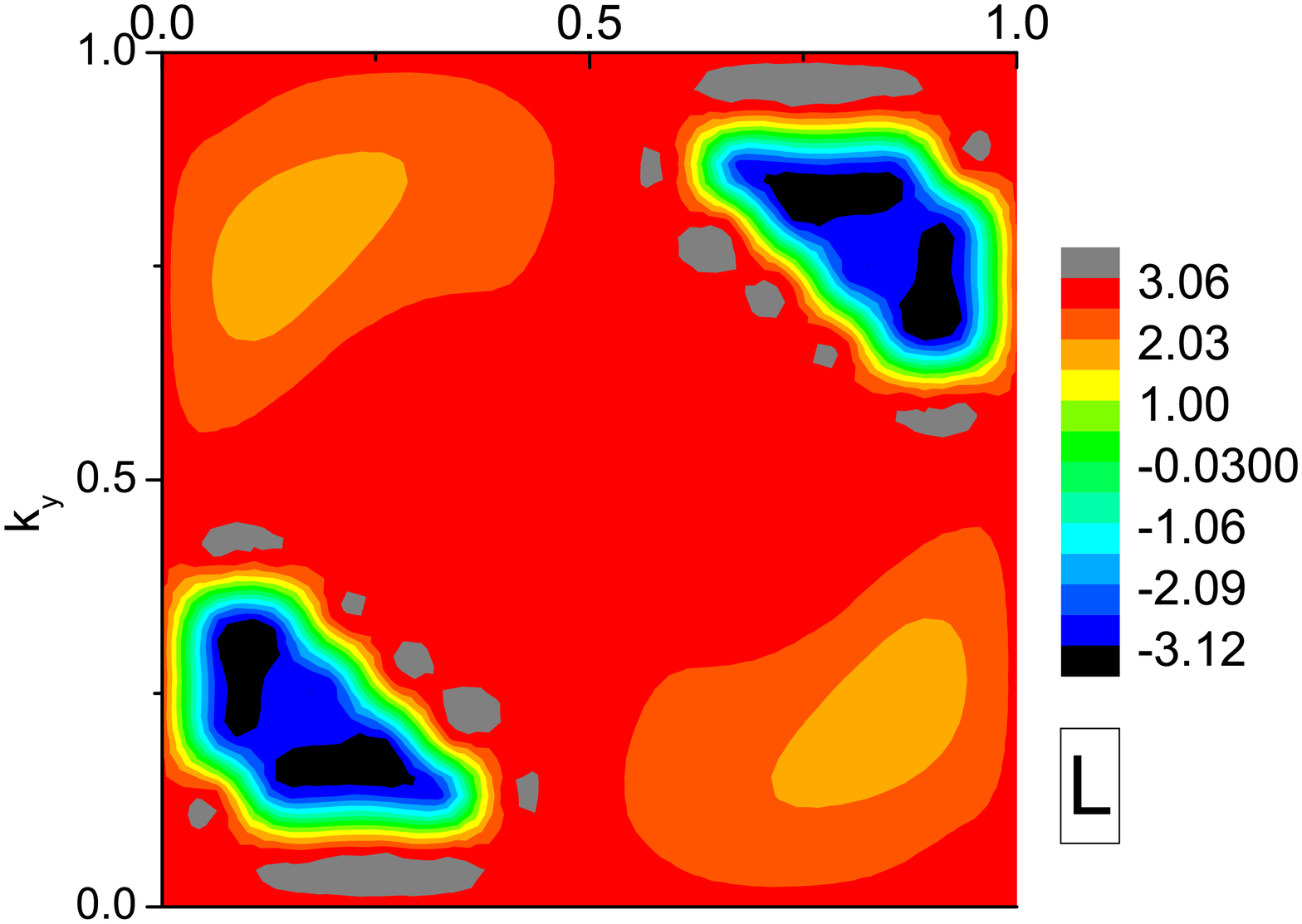,width=6cm}  \epsfig{file=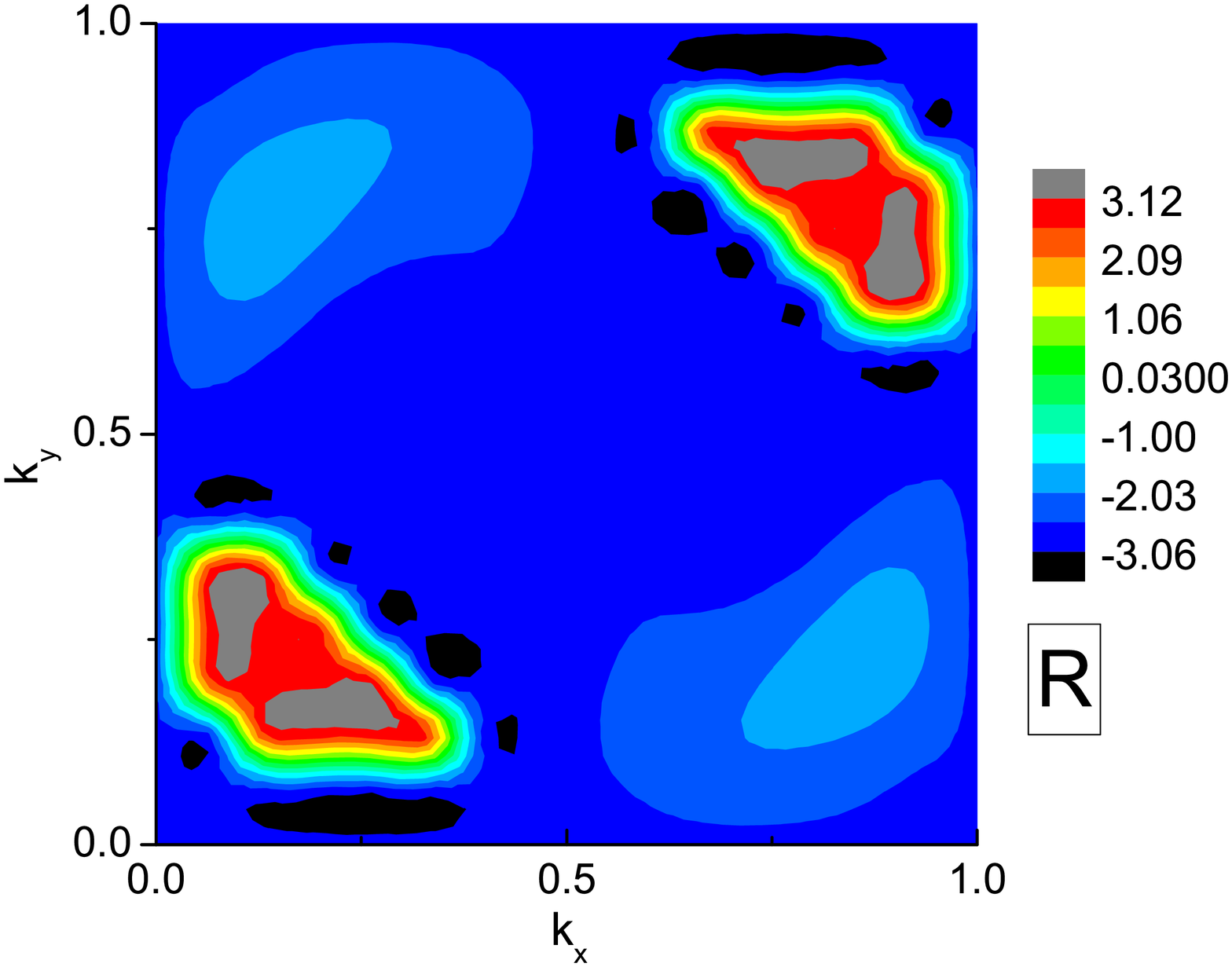,width=6cm}
   \caption{(Color online) Computed values of the local electronic Berry phase of FeSi as given by the function
            $\phi^{(\lambda=0)}(\mathbf{k}_{\bot})$ for both the right ($R$) and the left ($L$) handed enantiomorphs for  $\mathbf{k}_{\bot}$ in the $k_x , k_y$ plane. Spin-orbit coupling is included in these calculations, for which the ASW method was used. }
   \label{fig5}
   \end{center}
\end{figure}

The three components of the BP are obtained for FeSi using the value of $u(\rm {Fe})$ and $u(\rm {Si})$ given above; including spin-orbit coupling in the calculations we obtain $\Phi^{(\lambda=0)}=2.0$ for one of the three components in the ASW-method. The index $\lambda=0$ marks one point of the adiabatic path that is to be continued subsequently. If all the position vectors in the unit cell are inverted, \textit{i.e.} if the right-handed structure is used, $\Phi^{(\lambda=0)}=-2.0$. The Berry phases for RuSi and OsSi have approximately the same values, being $\Phi^{(\lambda=0)}\pm 1.7$. Since the BP is obtained from the imaginary part of the logarithm its value  must be in the the interval from $-\pi$ to $\pi$, hence it is $mod~ 2\pi$.  Thus the BP is non-zero in the three compounds. It is a vector parallel to the [111]-direction and its sign reflects the handedness of the crystal. It thus is a measure of the topology of the crystal. This should be seen in the context of symmetry-determined topological properties of insulators \cite{niu}, but should not be confused with topological insulators \cite{zhang,ando}. Furthermore, we  remark  that SOC is not necessary to obtain these results; it was included initially to extend the calculations to the heavier compound OsSi, however, a better reason  will emerge below. Results obtained for the Berry phase of FeSi by means of the VASP code \cite{vasp} are in good agreement with the ASW-results. In Fig.~\ref{fig5} we show the local electronic Berry phase for FeSi, given by
 $\phi^{(\lambda=0)}(\mathbf{k}_{\bot})$ for  $\mathbf{k}_{\bot}$ in the $k_x, k_y$ plane. One sees that it varies within $\pm \pi$, and one is the mirror image of the other, with respect to the signs.

Can the Berry phase be measured? Under certain well defined conditions it gives the macroscopic electric polarization, which thus emerges as a quantum-mechanical effect \cite{resta,resta1}. The condition is an available adiabatic path which connects two states labelled by $\lambda=0$ and $\lambda=1$. In this case $ P=-(e/\pi a^2)(\Phi^{(1)}-\Phi^{(0)})$ is the value of the polarization in a direction normal to the area $A$, $e$ being the electron charge and $a$  the lattice constant. Furthermore, the band structure has to be gapped at all points of the adiabatic path.

\begin{figure}[htb]
   \begin{center}
   \epsfig{file=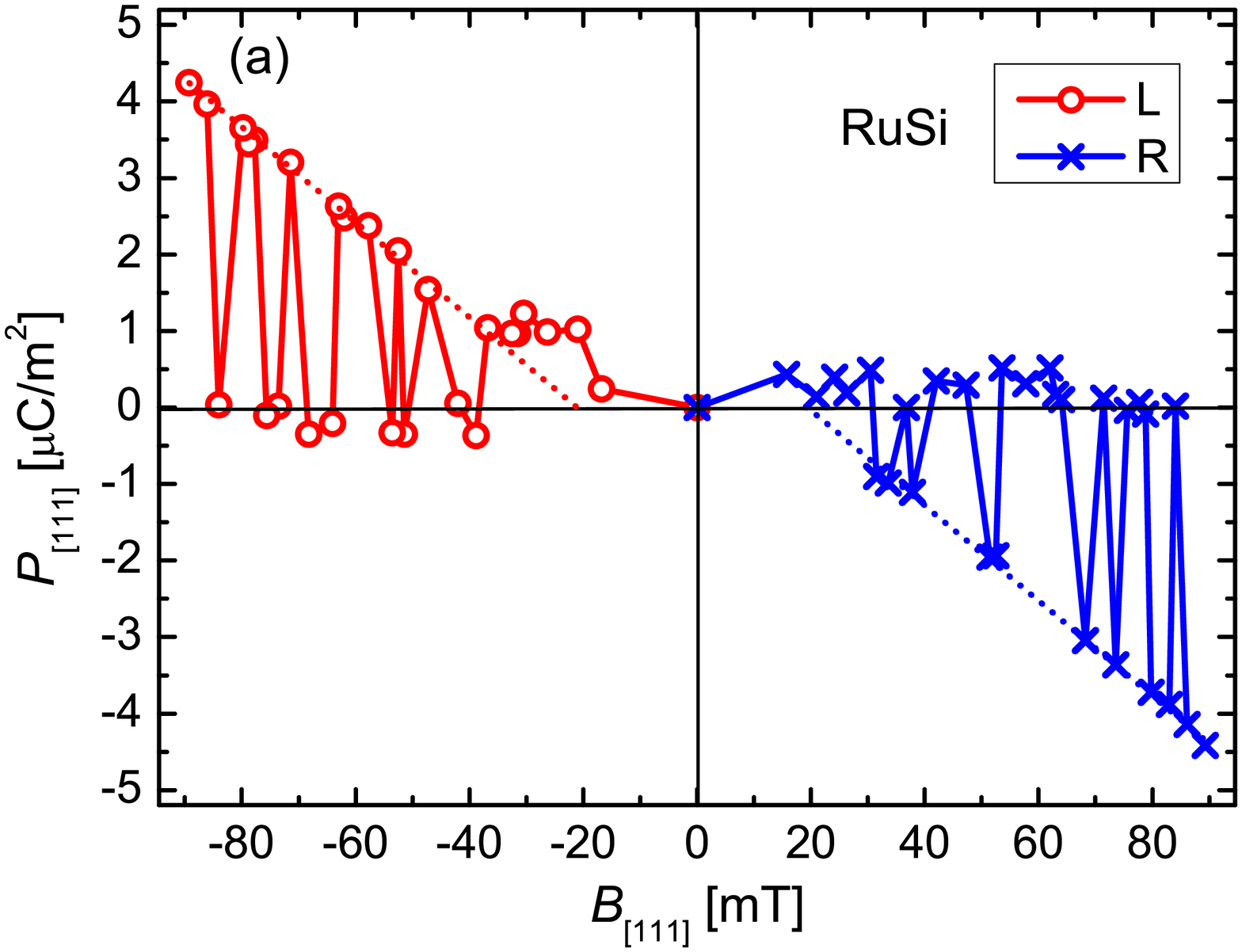,width=6cm}
   \epsfig{file=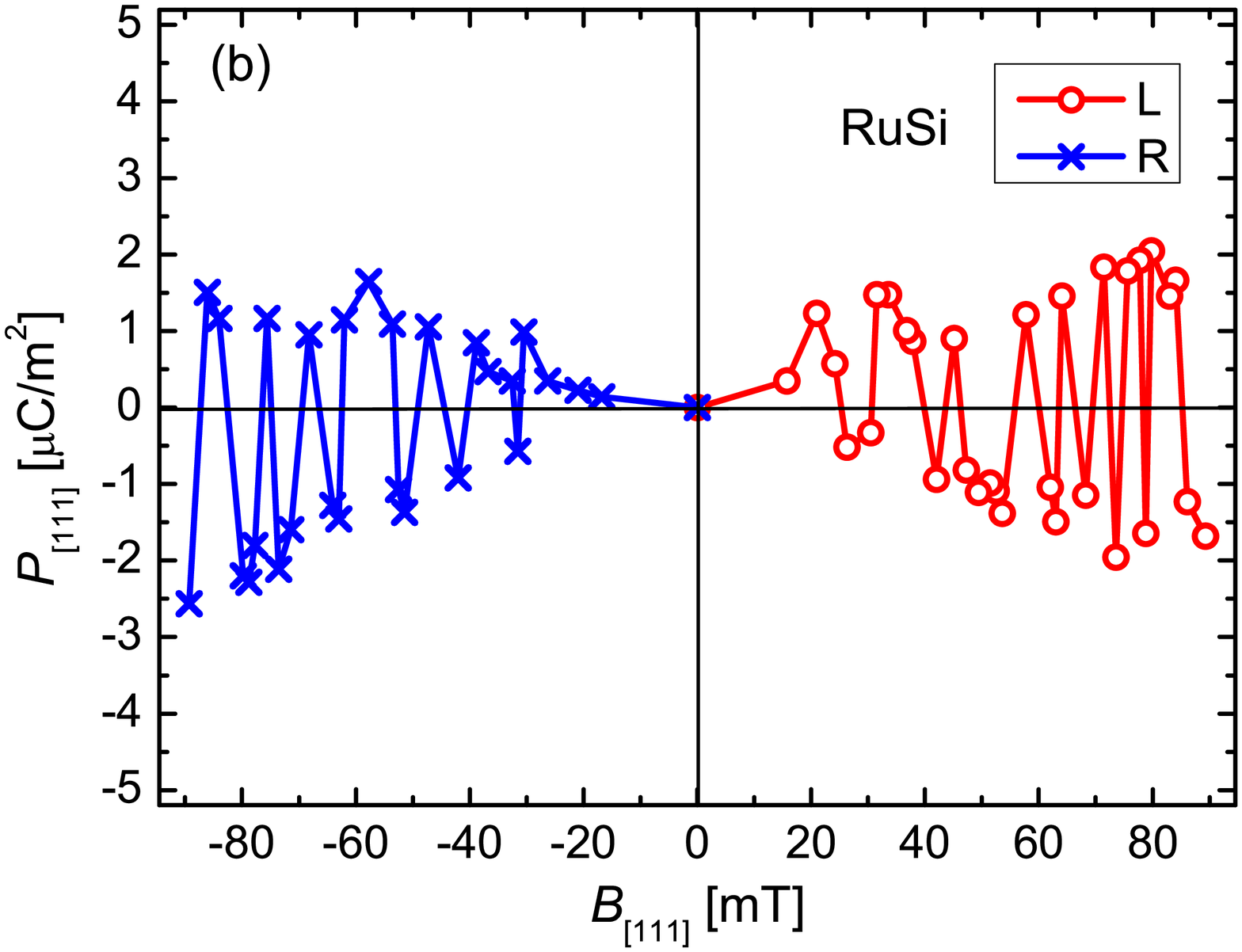,width=6cm}   
   \caption{(Color online) The polarization of RuSi in the [111] direction, $P_{[111]}$,  as a function of the magnetic field in the [111] direction, $B_{[111]}$. Part (a): directions of the magnetic field and the off-set polarization, $P_{OS}$, are the same. Part (b) directions of the magnetic field and $P_{OS}$   are opposite. The labels $L$ and $R$ stand for left- and right-handed chiral symmetry, corresponding with data points marked as circles and crosses, respectively.  The lightly dotted lines in part (a), red for negative fields and blue for positive fields, are a guide to the eye to see the slopes discussed in the text.}
   \label{figP}
   \end{center}
   \end{figure}
   
\begin{figure}[htb]
   \begin{center}
   \epsfig{file=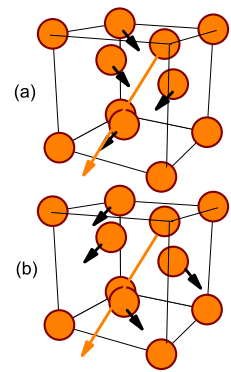,width=3cm} 
   \caption{(Color online) Sketch of the induced non-collinear moments showing only the metal atoms for the left-handed symmetry in a negative magnetic field. Large arrows indicate the direction of the off-set polarization, $P_{OS}$, small arrows the induced moments omitting those obtained by translations. Part (a) for the large polarization,  part (b) for the zero or near zero polarization. The $x$ and $y$-components of the polarization separately add up to zero, the $z$- component is in the direction of the magnetic field, the moments being at most of the order of $ 5 \cdot 10^{-4}\mu_B$ .  } 
   \label{fig8}
   \end{center}
\end{figure}

In trying to construct a plausible adiabatic path one could attempt to trace out  the continuous chirality measure discussed in connection with Fig.~\ref{fig11}.  The total energy changes, however, signal unphysical states for relevant deviations from the equilibrium making this approach unrealistic. Next, considering changes of the symmetry we note that the group $P2_13$ is non-polar. By breaking part of the symmetry it can be made polar. Thus mechanical deformations will supply  adiabatic paths, leading to piezoelectric effects. 

Following a different route, however, by applying a magnetic field, we notice that the spin-orbit interaction breaks the three-fold symmetry if the standard quantization axis (c-axis) is left intact. The further application of a magnetic field in the [111]-direction merely breaks the time-reversal symmetry. Technically the magnetic field is mimicked by a small Zeeman term, $\overrightarrow{\sigma}\cdot \bf {B}$, where $\overrightarrow{\sigma}$  is the Pauli spin vector. Because of SOC it induces both a small orbital and spin response, the latter being negligible in the Si atoms. In the metal atoms both effects cause a strong non-collinear moment configuration, which lead to symmetry properties that are described by spin-space groups \cite{brink,sandratskii}. This is so even though the individual moments are rather small being of the order of $ 5 \cdot 10^{-4}\mu_B$. Our central result is given in Fig.~\ref{figP},  which shows the electric polarization as a function of the magnetic field for left- and right-handed RuSi. For FeSi the polarization looks very similar but the amplitude is larger by about 20\%, for OsSi the results are of the same order as for RuSi. Calling the Berry phase converted to polarization \textit{off-set polarization}, defined by $ P_{OS}=-(e/\pi a^2)\Phi^{(0)}$, we show our results in two parts; in Fig.~\ref{figP} (a) the directions of the magnetic field and the off-set polarization are the same, in Fig.~\ref{figP} (b) the directions of the magnetic field and $P_{OS}$ are opposite.   

 A striking property of the computed data collected in Fig.~\ref{figP} are the oscillations which break a rather regular behavior. This behavior is different depending on the sign of the magnetic field with respect to the off-set polarization as is clearly seen comparing Fig.~\ref{figP} (a) with Fig.~\ref{figP} (b). There is no symmetry argument that requires the two figures to be the same.  To support the numerics of the  results the convergence was tested by doubling the number of k-points, which changed the computed polarizations by less than 1\%. 

The origin of the oscillations can be traced back to two different configurations of the non-collinear moments that are sketched in Fig.~\ref{fig8} for the left-handed crystal in a negative magnetic field. These configurations behave like \textit{nearly-degenerate states} that are formed by converging the electronic structure in the magnetic field. These states possess different polarizations; the one shown  in Fig.~\ref{fig8} (a) is large, the one in Fig.~\ref{fig8}~(b) is near zero, corresponding to the negative field values in Fig.~\ref{figP}~(a). The other cases are explained in an analogous way, except that the near-zero polarizations are larger in the case of Fig.~\ref{figP}~(b). In fact, one can manipulate the out-come of a calculation by constraining the states and thus \textit{switch} from one value to the other. Therefore, one should not expect to find any periodicity in the oscillations.     

For magnetic fields in the interval from about $-20$ mT to 20 mT the polarization scales roughly with the square of the field. For fields smaller than $-20$ mT the $L$-polarization has a negative slope as does the $R$-polarization for fields larger than 20 mT. Both slopes are indicated in the Fig.~\ref{figP} (a) by lightly doted lines. Both have the same value, a value which is also found for FeSi and OsSi. So there is approximately a single number for the slope of the three compounds, which we may convert to $cgs$ units obtaining
 $|dP_{[111]}/dB_{[111]}|\approx1.8\cdot10^{-3}$.

To appreciate the order of magnitude we compare this number with the size of the
topological magneto-electric (TME) effect \cite{qi}, where $ d {\bf P}/d {\bf B}=(n+\frac{1}{2})\frac{\alpha} {2\pi}$. Here $n$ is an integer and $\alpha$ is the fine-structure constant \cite{qi}. With $n=1$ this gives  $ d {\bf P}/d {\bf B}=1.74\cdot10^{-3}$, which is rather close to our numerical value.  We emphasize, however, that we do not imply our results are due to the TME. We add that we computed the first Chern number (or rather invariant \cite{resta1}) by means of the Berry curvature \cite{kubler2} with and without the magnetic field and find it within numerical accuracy to be zero.

It is important to compare these results with experimental data. They do not yet exist for the three compounds studied here, but recent work by Seki \textit{et al.} \cite{seki} describes the discovery of magnetoelectric skyrmions in an insulating chiral magnet Cu$_2$OSeO$_3$ which belongs to the same space group as our crystals, namely $P2_13.$  Seki \textit{et al.} observe a magnetization dependent electric polarization when applying a magnetic field in the [111] direction. Although different in many details, the polarization is of the same order of magnitude as our computed values and occurs for roughly the same magnetic-field strengths as ours. These authors argue that the various magnetic phases occurring in Cu$_2$OSeO$_3$ magnetically induce a non-zero electric polarization with varying sign and magnitude. What emerges from our calculations is that a property of the chiral crystal structure, the non-zero Berry phase, exists independent of any magnetic order, but, indeed, only becomes observable through internal or external magnetic fields. It may also indicate the existence of possible topological crystalline insulators protected by the chiral crystalline symmetry, similar to what was discussed in recent work \cite{niu,liu}. 

The question remains whether a single crystal of any of the compounds discussed here will be purely left or right handed. Neutron-diffraction experiments can clarify the question of enantiomorphs as the work of Grigoriev \textit{et al.}~\cite{grigoriev} demonstrates for magnetic B20 compounds. 
Inspecting the electronic polarization described in Fig. \ref{figP}, one could indeed argue that the magneto-electric effect should be observable in pure enantiomorphs of  FeSi, RuSi, and OsSi.
\bigskip
\begin{acknowledgments}    
We thank Gerhard~H.~Fecher for his critical discussions as well as for his help with the figures and Janos Kiss for participating in the  calculations.
\end{acknowledgments}

\end{document}